\documentclass[conference]{IEEEtran}
\IEEEoverridecommandlockouts
\usepackage[T1]{fontenc}
\usepackage{graphicx}
\usepackage{amsmath,amssymb}
\usepackage{booktabs}
\usepackage{array}
\usepackage{tabularx}
\newcolumntype{Y}{>{\raggedright\arraybackslash}X}
\newcolumntype{C}{>{\centering\arraybackslash}X}
\usepackage{enumitem}
\usepackage[table]{xcolor}
\usepackage{url}
\usepackage[hidelinks]{hyperref}
\usepackage{fancyhdr}
\newcommand{\claimbox}[1]{\begin{center}\fbox{\parbox{0.95\columnwidth}{\footnotesize #1}}\end{center}}

\fancypagestyle{firstpage}{%
  \fancyhf{}

  \fancyfoot[L]{\fontsize{8}{9.5}\selectfont%
    SC26 Workshops, November 15--20, 2026, Chicago, Illinois, USA\\%
    979-8-3195-1221-5/26/\$31.00 \textcopyright{}2026 IEEE}%
}

\begin{document}

\title{Privacy Foundations for Multi-Institutional\\ Scientific Artificial Intelligence}

\author{
\IEEEauthorblockN{Olivera Kotevska}
\IEEEauthorblockA{\textit{Oak Ridge National Laboratory}\\
Oak Ridge, TN, USA\\
kotevskao@ornl.gov}
\and
\IEEEauthorblockN{Sumit Jha}
\IEEEauthorblockA{\textit{University of Florida}\\
Gainesville, FL, USA\\
sumit.jha@ufl.edu}
\and
\IEEEauthorblockN{Aurélien Bellet}
\IEEEauthorblockA{\textit{Inria}\\
Montpellier, France\\
aurelien.bellet@inria.fr}
\and
\IEEEauthorblockN{Rui Hu}
\IEEEauthorblockA{\textit{University of Nevada}\\
Reno, NV, USA\\
ruihu@unr.edu}
\and
\IEEEauthorblockN{Nathaniel D. Bastian}
\IEEEauthorblockA{\textit{Johns Hopkins University}\\
Baltimore, MD, USA\\
ndbastian@jhu.edu}
\and
\IEEEauthorblockN{Rafael Ferreira da Silva}
\IEEEauthorblockA{\textit{Oak Ridge National Laboratory}\\
Oak Ridge, TN, USA\\
silvarf@ornl.gov}
\and
\IEEEauthorblockN{Ravi Madduri}
\IEEEauthorblockA{\textit{Argonne National Laboratory}\\
Lemont, IL, USA\\
madduri@anl.gov}
\and
\IEEEauthorblockN{Kibaek Kim}
\IEEEauthorblockA{\textit{Argonne National Laboratory}\\
Lemont, IL, USA\\
kimk@anl.gov}
}
\maketitle
\thispagestyle{firstpage} 

\begin{abstract}
Scientific artificial intelligence (AI), spanning foundation models (FMs) to federated data-analysis
pipelines, is becoming shared infrastructure across national laboratories,
universities, hospitals, and industrial partners. This collaboration creates
privacy risks whose natural unit is often an institution's participation,
research strategy, or technical capability rather than a single record.
Differential privacy (DP), federated learning (FL), secure computation, trusted
execution, and provenance each protect parts of the stack, but their guarantees
rarely compose across mixed-trust institutions, access tiers, and autonomous
agents. This perspective recasts privacy for scientific AI as an assurance problem
defined by six elements: \textit{protected asset}, \textit{observer}, \textit{channel}, \textit{permitted
disclosure}, \textit{guarantee}, and \textit{evidence}. We demonstrate the framing through a claim
register for a composite cross-institutional scenario and use it to
assess the model lifecycle. Two of the resulting gaps are specific to leadership-class facilities: per-round timing and scheduler metadata let peers infer an institution's participation, and instrument-attached control loops leak research strategy through timing and contention on shared accelerators. We identify six research priorities:
\textit{institution-level guarantees}, \textit{agent-communication privacy}, \textit{cross-tier
information flow}, \textit{privacy-compatible reproducibility}, \textit{leadership-scale
accounting}, and \textit{instrument side channels}. The contribution is a common form for stating, comparing,
and auditing claims whose guarantees otherwise remain fragmented across the scientific AI stack.
\end{abstract}


\section{Privacy Becomes Infrastructure}\label{sec:intro}
Scientific AI is moving from stand-alone research artifacts to shared
infrastructure, and foundation models (FMs) are its most visible case. Pre-trained
models now support molecular design, climate modeling, biomedical analysis, and
materials discovery, while autonomous laboratories increasingly connect
machine learning systems to simulation, instruments, and robotic
experimentation~\cite{kotevska2025privacy}. The data and
facilities these systems require are rarely controlled by one organization; they
are distributed across national laboratories, universities, hospitals,
observatories, and industrial partners, under incompatible obligations
concerning consent, intellectual property, export control, and reproducibility.
The US Genesis Mission makes the trajectory explicit, directing the integration
of supercomputing, domain FMs, AI agents, and autonomous experimentation within a
secure platform spanning classification, privacy, and export-control
regimes~\cite{genesis2025}. Cross-facility federated learning (FL), which trains
without sharing raw data, already shows that large-model training can be
orchestrated across heterogeneous leadership-class
systems~\cite{kotevska2026scalable, li2026scalable}.

\begin{figure}[t]\centering
\includegraphics[width=0.9\columnwidth,height=0.22\textheight,keepaspectratio]{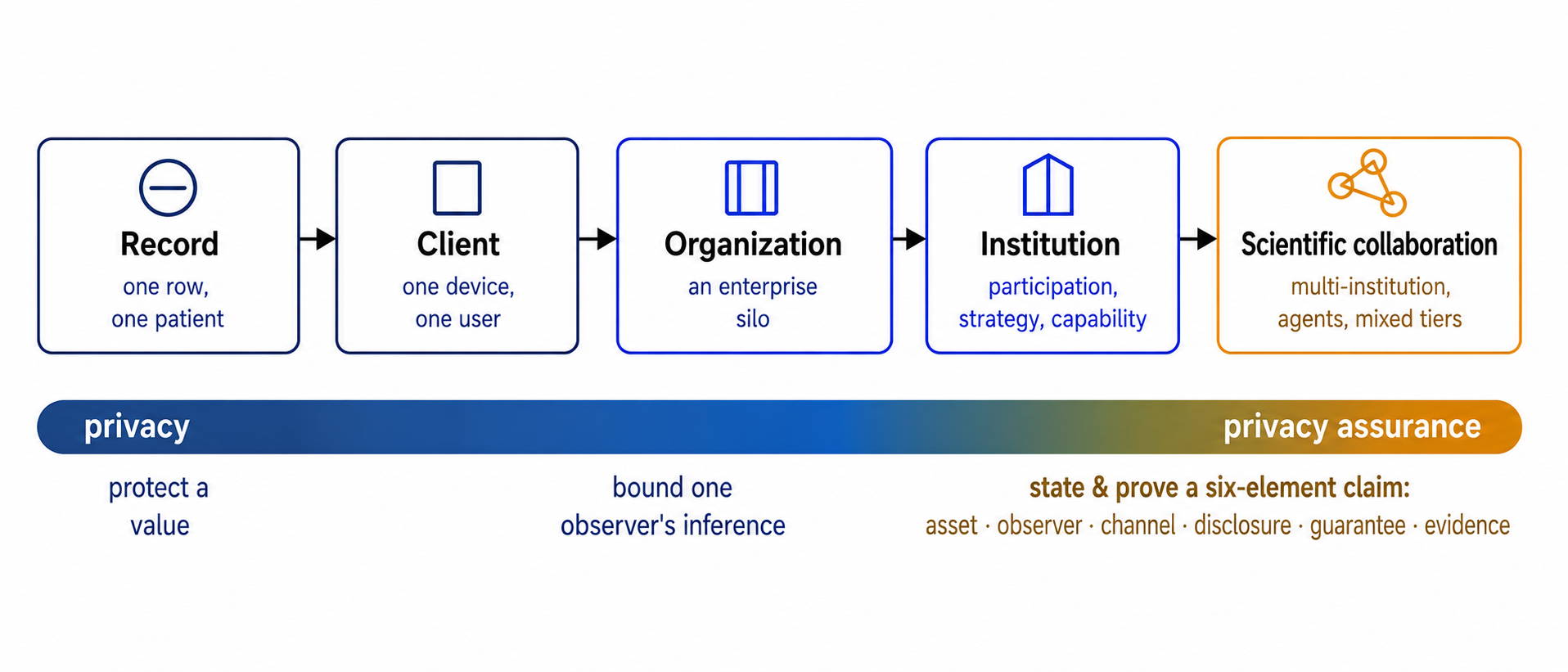}\vspace{-15pt}
\caption{From record- and client-level privacy to institution-level privacy
assurance. As the protected unit grows, a claim must guarantee and \emph{prove}
more; multi-institutional scientific AI sits at the right, where guarantees must
compose across institutions, tiers, and agents.}
\label{fig:evolution} \vspace{-15pt}
\end{figure}

Enterprise cross-silo learning is a poor model for these arrangements
(see Fig.~\ref{fig:evolution}). Scientific facilities are shared user
resources; the same instrument or supercomputer may serve projects governed by
different agreements, and one institution can simultaneously be a collaborator,
an infrastructure operator, a model consumer, and an audit authority. The value of a dataset may lie in a rare physical regime rather than in a count
of records, which makes common privacy-utility intuitions unreliable. Scientific
claims also outlive the training run, so controls must persist across
organizational and temporal boundaries. Facility-oriented high-performance computing (HPC) security guidance characterizes
system architecture, threats, and posture at the level of a facility and its
operator~\cite{nist800223}; the institution-level confidentiality questions we
raise here are complementary and cut across that view
(see Table~\ref{tab:comparison}).

Privacy-preserving machine learning supplies essential components but not a
complete architecture~\cite{kim-kotevska2025privacy}. FL keeps raw data at its source~\cite{kairouz2021}, yet shared gradients may
leak training examples~\cite{zhu2019}. DP~\cite{DworkR14} quantifies
protection against such leakage, but its guarantees are defined over records,
users, or clients~\cite{abadi2016, xu2026optimal, kotevska2026dptwolevel} and do not hide
an institution's participation, strategy, or capability. Secure
aggregation~\cite{bonawitz2017secureagg}, multiparty computation
(MPC)~\cite{yao1982gc}, and trusted execution environments
(TEEs)~\cite{costan2016sgx} restrict what operators observe during computation,
not what a trained model or agent may later disclose. Scientific
AI does not lack privacy techniques. The available guarantees protect different
assets, against different observers, through different channels, and do not
automatically compose (Table~\ref{tab:comparison}). Multi-institutional
scientific AI therefore needs a \emph{privacy assurance stack} in which
governance, statistical guarantees, cryptographic controls, information-flow
policies, trusted hardware, and auditable evidence are aligned to an explicit
threat model (see Fig.~\ref{fig:assurance}), preserving collaboration and
scientific validity rather than treating data non-disclosure as the only
objective.

\textbf{Perspective and contributions.} Although FMs motivate the discussion, the
framework applies to multi-institutional scientific AI broadly, including
federated analytics and autonomous data-science pipelines. It makes three
contributions. \textit{First}, it argues that the protected unit often extends
beyond individual records to institutional participation, research strategy,
technical capability, and tier-restricted knowledge. \textit{Second}, it
introduces a six-element assurance claim as a common form for stating and
comparing privacy claims across heterogeneous mechanisms. \textit{Third}, it
applies that framing through an illustrative claim register and lifecycle
analysis to expose compositional gaps across distributed training, model release,
autonomous agents, provenance, HPC infrastructure, and instruments.

\textbf{Composite illustrative scenario.} Throughout, four partners jointly train a multimodal FM: a national laboratory
operating a light source and a leadership-class computer, a university
contributing method development, a hospital contributing a clinical imaging
cohort, and an industrial partner contributing proprietary catalyst formulations.
They then deploy autonomous agents that plan experiments and schedule beamline
time across all four sites. It deliberately combines recurring characteristics of
scientific AI collaborations: institutional data ownership, shared
leadership-class computing, heterogeneous policy obligations, tiered access,
autonomous agents, instruments, and long-lived provenance requirements.

\section{The Unit of Harm Is Not Always a Record}\label{sec:unit}
The word \emph{privacy} is used broadly in scientific collaborations. We use it
as shorthand for \emph{controlled disclosure}, while keeping it distinct from
confidentiality (keeping a specific value secret), access control (who may read a
resource), and information-flow security (how information may move between
security domains once inside the system). In the running consortium, the hospital
wants statistical privacy for its cohort, the industrial partner wants
confidentiality for a formulation, and the shared model needs information-flow
control so that an open-tier user cannot recover restricted inputs. These are
different guarantees, not one.

Table~\ref{tab:assets} lists the assets scientific consortia add beyond the
record-level framing, each with its observation channel and the harm it carries.
Four are properties an institution holds: \emph{participation}, \emph{research
strategy}, \emph{capability}, and \emph{tier-restricted knowledge}, the last
admissible at one access level but not another across the open, controlled, and
classified tiers leadership-class facilities operate. Two more are systemic and
inferable from scheduler and orchestration metadata: \emph{resource posture} and
\emph{collaboration structure}. They correlate: a sequence of individually
innocuous agent messages can jointly reveal an institution's objective. 
These assets cannot be protected merely by scaling record-level privacy from individuals to institutions. An institutional contribution may contain millions of correlated observations, making the resulting group privacy guarantee too weak to be useful. Participation, strategy, and other institutional assets may also be inferred from aggregate behavior even when no individual record is exposed. 
The research
task is therefore to define neighboring worlds and leakage functions matched to the harm.

\begin{table}[t]
\caption{Institution-level privacy assets in scientific AI. The examples are
illustrative; the protected unit and resulting harm remain deployment-specific.}
\vspace{-0.6\baselineskip}
\label{tab:assets}
\centering\scriptsize
\renewcommand{\arraystretch}{1.04}
\setlength{\tabcolsep}{1.6pt}
\begin{tabularx}{\columnwidth}{@{}p{1.55cm}YYY@{}}
\toprule
\textbf{Asset} & \textbf{Example} & \textbf{Observation channel} & \textbf{Representative harm} \\
\midrule
Participation & Whether a hospital or beamline contributes in a given period & Round timing, model updates, public membership & Reveals cohort, program, or direction \\
Research strategy & Hypotheses, variables, or parameter regions & Agent messages, tool calls, job sequence & Loss of scientific priority or competitive advantage \\
Capability & Instrument precision, control policy, or model competence & Calibration data, outputs, telemetry, side channels & Reveals facility or industrial capability \\
Tier-restricted knowledge & Controlled result or restricted simulation condition & Outputs, adapters, retrieval, provenance & Cross-tier disclosure \\
Resource posture & Scale, urgency, or cadence of a campaign & Allocations, queue timing, checkpoints, accounting & Reveals operational intent \\
Collaboration structure & Which partners interact and when & Orchestration graph, communication metadata & Reveals program relationships \\
\bottomrule
\end{tabularx}
\end{table}

The adversary is frequently inside the collaboration. An honest-but-curious
institution may follow the protocol while analyzing every update it is authorized
to receive; privacy attacks by curious peers in decentralized learning show that this
threat is realistic~\cite{elmrini2024}. A competitively motivated participant may
adapt queries to infer a partner's unpublished results; an external attacker may
compromise the orchestrator; and a compromised agent may misuse legitimate
credentials following prompt injection or a supply-chain compromise. Classical
record-level attacks such as membership inference and training-data
extraction~\cite{shokri2017, carlini2021} remain relevant but target the wrong
unit. Dataset- and property-inference attacks~\cite{maini2024llmdatasetinference, ganju2018propertyinference}
move closer to the institutional question of whether a party's data or a global
property was learned, but the observation set that matters extends further, to
participation, strategy, and semantic agent traffic (Section~\ref{sec:autonomous}). We scope this
paper to confidentiality and controlled disclosure. Integrity, availability,
poisoning, model safety, fairness, and research ethics may share observers,
channels, and evidence with privacy claims, but they need different guarantees,
and calling all of them ``privacy'' obscures which property is claimed.

\subsection{Making the neighboring world precise}
\label{sec:neighboring}
The requirement can be stated in the Pufferfish
framework~\cite{kifer2014pufferfish}, which generalizes DP to arbitrary secrets
under an explicit data-generating model. Let $\Theta$ be a class of
distributions over consortium data and scheduler behavior. For institution $i$
and round set $R$, let $s_i^R$ denote the secret ``$i$ contributed in $R$'' and
$\neg s_i^R$ its complement. Write $V_O$ for the view an observer $O$ obtains
(Eq.~\ref{eq:view}). A deployment provides \emph{$\varepsilon$-participation
privacy against $O$} if for every $i$, every $R$, every $\theta\in\Theta$ with
$\Pr[s_i^R\mid\theta]>0$ and $\Pr[\neg s_i^R\mid\theta]>0$, and every view $v$,
\begin{equation}
\Pr[V_O = v \mid s_i^R,\theta] \;\le\; e^{\varepsilon}\,
\Pr[V_O = v \mid \neg s_i^R,\theta].
\label{eq:pufferfish}
\end{equation}
Record-level DP is the special case in which the secret pair concerns the
presence of one row and $V_O$ is the mechanism output alone. Three things change at the institution level. The secret is a property of a
participant rather than a row, so group privacy does not apply. $\Theta$ must
encode correlations between an institution's corpus and its scheduling behavior,
which makes the definition hard to instantiate rather than hard to state. And
$V_O$ ranges over released weights, per-round timing, and orchestration metadata,
not a single output. Substituting research strategy or capability for participation
yields the corresponding definitions. The in-scope observers are those named in
Table~\ref{tab:register}: open-tier users, colluding peers, the aggregation
operator, co-located tenants, and authorized auditors; physical access and
credential theft are out of scope. We state the target rather than claim it. No
deployment in Table~\ref{tab:register} meets it, and choosing $\Theta$ for a
scientific consortium remains open.

Because $V_O$ is a union (Eq.~\ref{eq:view}), an $\varepsilon$-guarantee for each
$V_{O,i}$ does not imply one for $V_O$ unless the constituent channels are
conditionally independent given $\theta$. That is the formal statement of the
composition failure in Section~\ref{sec:worked}.

\section{A Six-Element Privacy Assurance Claim}\label{sec:framework}
We represent a privacy assurance claim as
\begin{equation}
\mathcal{K}=(A,O,C,D,G,E),
\end{equation}
where $A$ is the protected asset, $O$ the observer, $C$ the observation
channel, $D$ the permitted disclosure, $G$ the claimed guarantee, and $E$ the
evidence. The form draws on established security-assurance, threat-modeling, and
information-flow traditions, but adapts them to scientific AI settings in which
institutions, access tiers, agents, HPC services, and instruments create
observation surfaces that cross mechanism boundaries. The novelty is therefore
not that assurance requires assets, threats, and evidence in the abstract; it is
the institution-level and lifecycle-spanning form of the claim and its use as a
common register across heterogeneous privacy mechanisms.

The form deliberately stays close to existing practice; its contribution is what
it adds for this setting. Data-flow privacy threat modeling
(LINDDUN~\cite{linddun}) and security threat modeling (STRIDE~\cite{stride})
enumerate threats over records and data-flow elements, but carry neither
institutional assets, agent channels, nor cross-tier composition. Run on the
scenario of Section~\ref{sec:intro}, LINDDUN produces linkability and
identifiability threats on the staging area and the update stream, recovering C1
and C2, but it has no asset category for whether an institution participated, so
C3 does not arise. STRIDE places information-disclosure threats at the trust
boundaries it recognizes, which are process and data-store boundaries, and draws
none at the instrument control loop, so C6 does not arise. The two claims the register
marks as gaps are exactly the two these frameworks have no place to put.
Assurance-case notations such as goal-structuring notation~\cite{gsn} organize
evidence into an argument while leaving the privacy claim itself unspecified. The six-element claim combines the argue-from-evidence stance of an
assurance case with the enumerate-the-surface stance of threat modeling,
instantiated at the institution level across the scientific-AI lifecycle, the
level at which the harms of Section~\ref{sec:unit} actually arise. The claim also
operationalizes \emph{contextual integrity}~\cite{nissenbaum2004ci}: its observer,
channel, and permitted-disclosure elements correspond to the recipient,
information type, and transmission principle of an appropriate-flow norm, lifted
from the individual to the institution.

A useful privacy claim answers six questions:
\textbf{(1)~What is protected?} the asset, from a patient record to an
institution's participation or an instrument capability. \textbf{(2)~From
whom?} the observer: an external querier, the aggregation service, a
collaborating institution, an authorized auditor, or a compromised agent.
\textbf{(3)~Through what channel?} raw data, gradients, checkpoints, adapted
weights, outputs, tool calls, messages, memory, timing, or logs.
\textbf{(4)~What disclosure is permitted?} the abstraction or statistic allowed
to cross the boundary, since useful collaboration is never total secrecy.
\textbf{(5)~What guarantee is claimed?} DP, simulation-based cryptographic
security, non-interference, an access-control policy, or a bounded attack-success
rate. \textbf{(6)~What evidence demonstrates it?} a privacy accountant, proof,
attestation, audit record, or benchmark result. Each element removes a
specific ambiguity: without an asset, it is unclear what is protected; without
an observer, against whom; without a channel, where observation occurs; without
permitted disclosure, useful collaboration cannot be distinguished from leakage;
without a guarantee, the statement is aspirational; and without evidence, it is
not testable. We refer to these elements by number, (1)--(6), throughout; they
index the columns of the claim register in Table~\ref{tab:register}, so any
element reference in the text maps directly to a column of the register.

\subsection{Design principles for institution-level assurance}
Applied well, the form follows a few principles. Name a specific institutional asset, not ``the data''. Bind each guarantee to a defined observer and observation set, and count model, agent, infrastructure, metadata, and instrument paths as channels. State what may be declassified, and keep evidence proportional to the claim. Review composed views across mechanisms, stages, and institutions, binding each claim to a versioned workflow and validity period. These make the register an assurance artifact rather than a static inventory of controls.

Omitting any element makes a claim hard to test and easy to overstate, because
the same mechanism is strong in one configuration and weak in another. Secure aggregation ensures no party learns more than the aggregate, yet a colluding subset can subtract its own inputs to sharpen inference about a remaining participant, and the released aggregate can still leak participation. Likewise, an audit log can prove compliance to a reviewer yet become a disclosure channel if copied into a general analytics system. A claim must therefore be bound to a versioned workflow and
observation model, not to a model name. The permitted-disclosure element, the one
most often left implicit, separates assurance from secrecy: it turns ``keep the
data private'' into a testable contract (``a peer agent may learn a feasible
experimental region but not the rationale that made it strategically
important''). Table~\ref{tab:comparison} reads the major controls through the six
questions: no row is inadequate in isolation, but none answers all six for the
multi-institutional setting.

\begin{table}[t]
\caption{Existing controls read through the six-element claim. $\bullet$ marks
an aspect represented natively; ``partial'' indicates that coverage depends on
configuration or composition; -- marks an aspect outside the primitive's native
guarantee. The comparison concerns each primitive in isolation, not a complete
composed architecture. The last row is the proposed framework, not a primitive:
its marks indicate which aspects a register entry must state, not that the
register enforces them.}
\vspace{-0.6\baselineskip}
\label{tab:comparison}
\centering\scriptsize
\setlength{\tabcolsep}{1.6pt}
\begin{tabularx}{\columnwidth}{@{}YCCCCC@{}}
\toprule
Control & Protected unit & Internal comm. & Cross-tier flow & Agent channels & Lifecycle compos. \\
\midrule
Differential privacy    & record & -- & -- & -- & partial \\
Federated learning      & location & -- & -- & -- & -- \\
Secure aggregation      & update & -- & -- & -- & -- \\
Trusted execution       & compute & -- & -- & -- & -- \\
Info-flow control       & domain & partial & $\bullet$ & -- & partial \\
FM privacy frameworks   & mixed & -- & partial & -- & partial \\
\rowcolor{gray!15}
\textbf{Six-element register (framework)} & \textbf{institution} & $\bullet$ & $\bullet$ & $\bullet$ & $\bullet$ \\
\bottomrule
\end{tabularx}
\end{table}

\claimbox{\textbf{What this work does \emph{not} claim.} We propose no new DP algorithm, FL protocol, or replacement for cryptography or information-flow control. We propose an assurance framework connecting them: a common form for institution-level claims, a way to expose where guarantees fail to compose across the lifecycle, and a basis for specifying evidence that would support or falsify a claim.}

\begin{figure*}[t]\centering
\includegraphics[width=0.96\textwidth,height=0.21\textheight,keepaspectratio,trim=0 130 0 24,clip]{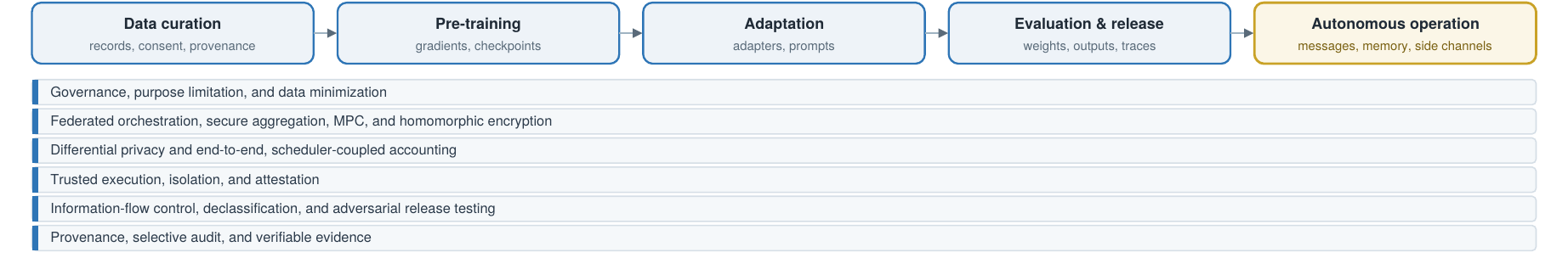}
\caption{A privacy assurance stack for multi-institutional scientific AI.
Observable channels change across the lifecycle, so no single primitive protects
the whole system; controls are layered from governance to provenance.}
\label{fig:assurance}
\end{figure*}

\section{A Worked Example: An Assurance Claim Register}\label{sec:worked}
Table~\ref{tab:register}
instantiates it as an illustrative \emph{claim register} for the composite
scenario: each row (labeled C1--C7) is one six-element confidentiality claim. We envision such a
register as an assurance artifact that a facility or consortium could maintain
for each deployment. Its purpose is to show how the framework exposes gaps that a mechanism-by-mechanism
review may overlook. Review reduces to three checks: every institution-level
asset has a claim; every claim names all six elements; and every ``none'' or ``--''
is recorded as an explicit residual risk rather than left as an oversight.

\begin{table*}[t]
\caption{Claim register for the consortium FM. Each row is a
six-element confidentiality claim. Cells marked \textbf{none}/\textbf{--}
(claims C3 and C6) are the deployment's explicit institution-level gaps: they
pass a per-mechanism control review yet leave the asset unprotected.}
\vspace{-0.6\baselineskip}
\label{tab:register}
\hyphenpenalty=10000\exhyphenpenalty=10000
\centering\scriptsize
\renewcommand{\arraystretch}{1.04}
\setlength{\tabcolsep}{1.8pt}
\begin{tabularx}{\textwidth}{@{}p{0.82cm}YYYYYY@{}}
\toprule
\textbf{Claim} & \textbf{(1) Protected asset} & \textbf{(2) Observer} & \textbf{(3) Channel} & \textbf{(4) Permitted disclosure} & \textbf{(5) Guarantee} & \textbf{(6) Evidence} \\
\midrule
C1 Data at rest & Hospital cohort records & Facility operator, other tenants & Shared storage, staging area & None (raw data stays in enclave) & Access control, encryption, TEE isolation & Attestation, access logs \\
C2 Aggregation & Hospital update (implies cohort) & Honest-but-curious aggregation server & Gradients, model deltas & Aggregate mean only & Secure aggregation (semi-honest) & Protocol proof, config attestation \\
\rowcolor{gray!12}
C3 Participation & \emph{Whether hospital participates} & Open-tier model users \emph{plus colluding peers} & Released weights, per-round timing & Public consortium membership, not per-round role & \textbf{Institution-level DP (target; none deployed)} & \textbf{Participation-inference advantage $\le\delta$ (target)} \\
C4 Release & Partner formulation space & Open-tier queries & Model outputs, adapter weights & Aggregate structure--property trends & Information-flow control, adversarial release testing & Tiered release certificate, red-team report \\
C5 Agent message & University hypothesis, target region & Compromised or colluding agent & Inter-agent messages, tool args, memory & Feasible region, not strategic rationale & Task-conditioned abstraction, policy filter (empirical) & Agent-leakage benchmark (partial) \\
\rowcolor{gray!12}
C6 Instrument channel & Beamline cadence (implies strategy) & Co-located tenant, electromagnetic observer & Timing, power, cache, control-loop latency & None & \textbf{Constant-time loop, jitter (often absent)} & \textbf{Side-channel trace audit (missing)} \\
C7 Audit & Full workflow provenance & External reviewer vs.\ authorized auditor & Provenance store, replay package & Filtered replay to auditor; proof to reviewer & Selective disclosure, zero-knowledge compliance proof & Signed manifest, audit record \\
\bottomrule
\end{tabularx}
\end{table*}

\subsection{Where individually sound claims fail to compose}
Claims C1, C2, and C4 each hold in isolation: raw records never leave the
enclave, the hospital's update is hidden from a semi-honest server, and the
released model passes information-flow and release testing. A per-mechanism
review signs off on all three.
Yet C3, \emph{institutional participation}, illustrates a plausible composition
failure. An open-tier user of the released model (C4's channel), combined with
public consortium membership and observable per-round aggregation timing (C2's
residual metadata), may let a colluding peer infer that a hospital of a given
cohort type participated in a given period. No single mechanism is violated. The leak is visible only because C3 is stated as its own claim, with an observer
set spanning C2's peers and C4's users. That is the case for treating
participation as a register entry of its own: composition is where assurance
breaks.

The channel C3 depends on has been measured. In federated fine-tuning on an
exascale system with up to 96 concurrent clients, per-round communication time
rises from roughly 33~s to 275~s as clients are added, and that figure reflects
synchronization and barrier waiting rather than transfer, since payload sizes are
unchanged~\cite{kotevska2026scalable}. Round completion time
therefore varies with who participates and when, at a magnitude visible to any
observer able to time rounds. This is not a participation-inference attack, which
is the evidence C3 still lacks, but it establishes that the channel is real at
leadership scale rather than hypothetical.

The converse question is when claims \emph{do} compose. Writing $V_O$ for the
combined view of Eq.~\ref{eq:view}, the following are sufficient for C2, C4, and
C5 to compose without opening a path none of them admits individually:
(i) their observer views are disjoint, or their union is analyzed explicitly
under Eq.~\ref{eq:view}; (ii) the residual metadata C2 leaves, round timing and
participation-set size, is either bounded or included in C4's release-risk check;
(iii) C4's declassification is monotone with respect to C5's task-conditioned
abstraction, so an agent receives no more than the released model already
permits; and (iv) no randomness or identifier is shared across the three
mechanisms. We state these as conjectured sufficient conditions. Proving them,
and finding the weakest set that suffices, is open; the deployment in
Table~\ref{tab:register} violates (ii).

\subsection{Where evidence is weaker than it looks}
\label{sec:weak}
The register also disciplines element~(6). C1 relies on TEE attestation, which
evidences that specified code \emph{ran}, not that it enforced the correct
disclosure policy for C4 or C5~\cite{zhu2024cc}; treating attestation as evidence
for a disclosure claim silently upgrades a weaker guarantee. C5's evidence is an
agent-leakage benchmark score~\cite{elyagoubi2026}, which bounds observed leakage
on a test distribution, not a worst-case adversary. This asymmetry is intrinsic:
privacy auditing can \emph{falsify} a DP claim by exhibiting an attack whose
success exceeds the stated guarantee, but it cannot \emph{certify} one, since a
stronger attack may always exist~\cite{cebere2025, cebere2026zeroauditing}; this
limit is acute for FMs, where retraining-based audits are
infeasible. Naming the evidence artifact precisely keeps the register falsifiable
and prevents a strong-sounding mechanism from masquerading as a strong guarantee.

\subsection{The channel that model-level analysis misses}
C6 is the claim a purely data- and model-centric review omits. Even with every
data and model channel sealed, the \emph{cadence} of beamline jobs and the
control-loop timing of an instrument-attached model leak research strategy
through timing, power, and cache contention~\cite{weiss2024, gao2025}. Shared HPC accelerators make this concrete: covert and side channels across co-located tenants have been demonstrated on the very multi-GPU machines scientific AI runs on~\cite{dutta2023spygpu}. This
channel crosses the instrument-to-compute boundary, so it is easy to leave off a
system-level review entirely, which is why it belongs in an asset-indexed
register, and where hardware-security methods meet institution-level assets.

\section{Using the Claim Register}\label{sec:procedure}
A consortium can apply the framework as a repeatable review procedure. Enumerate
the assets of Table~\ref{tab:assets}, the observers and authorized roles, and the
channels across data, training, adaptation, release, autonomous operation, HPC
services, instruments, and provenance. State the disclosure permitted through each
channel, attach a guarantee, and name the evidence artifact and workflow version.
Then inspect cross-channel and cross-stage composition, and record incomplete
claims as explicit residual risks. The
relevant HPC artifacts include scheduler and accounting logs, job and queue
timing, checkpoint metadata, parallel-file-system access, interconnect traffic,
accelerator telemetry, allocation patterns, instrument-control timing, workflow
provenance, and agent tool calls to facility services.

\subsection{Completeness beyond individual claims}
Completeness has a first-principles limit. Over an open-ended channel set it is
not decidable: an observation surface is not enumerable in advance, and a channel
nobody has named cannot be shown absent. What a register offers is completeness
\emph{relative to} an enumerated set of observers and channels, which is why the
procedure above fixes that set explicitly and records anything outside it as
residual risk. Automated assistance can widen the enumeration, deriving candidate
channels from scheduler logs, tool manifests, and dataflow graphs, and flagging
claims whose observer sets intersect; it cannot certify that the enumeration is
closed. C6 is the standing example: the instrument channel was absent from
model-centric reviews not because a test failed but because no one had listed the
channel.

A well-formed tuple is necessary but not sufficient for system-level assurance.
Beyond \emph{element completeness} (each claim names all six elements),
\emph{asset coverage} (every institutional asset appears in some claim), and
\emph{observer--channel coverage} (the relevant combinations of who observes
what), \emph{compositional completeness} requires analyzing views combined across
claims, stages, institutions, or time. For observer $O$, let
\begin{equation}
V_O = \bigcup_{i\in I_O} V_{O,i},
\label{eq:view}
\end{equation}
where 
\(I_O\) indexes the claims or channels observable to \(O\), and \(V_{O,i}\) is the corresponding view.
A deployment
may satisfy every claim under its individual view while failing under the
combined view $V_O$. The claim register therefore records cross-claim dependencies and residual
composition risks alongside isolated mechanisms.

Scale matters here. The register itself grows roughly as the product of assets
and observers, which is linear and manageable: the consortium of
Section~\ref{sec:intro} yields tens of claims, not thousands. The cost is in
composition, since $V_O$ ranges over subsets of the claims observable to $O$ and
analyzing all subsets is exponential in $|I_O|$. Two things keep this tractable.
Observers collapse into a small number of roles, peer, operator, open-tier user,
auditor, and co-located tenant, so review is per role rather than per party; and
only claims sharing a channel or an observer need joint analysis, which leaves
$I_O$ sparse. Choosing which combinations to analyze in larger federations, and
doing so with tool support, is open.

\section{Evidence and Assurance Maintenance}\label{sec:evidence}
The sixth element, evidence, turns a privacy statement into an assurance claim,
and is necessarily claim-relative: an artifact is meaningful only with respect to
a specified asset, observer, channel, permitted disclosure, and guarantee.
Section~\ref{sec:weak} showed two artifacts read as stronger than they are.

Table~\ref{tab:evidence} organizes evidence into six complementary classes. They are complementary rather than ordered: a strong assurance case commonly needs
several at once. Formal evidence can establish a protocol property under stated
assumptions, while operational evidence is still needed to show that the deployed
configuration, software version, and observation surface match those assumptions.
Adversarial evidence can expose violations that a proof model omitted, while
provenance binds the result to the workflow that was actually run.

\begin{table}[t]
\caption{Evidence classes for scientific-AI privacy assurance. No single class
is sufficient for every claim; the required combination depends on the guarantee
and observer model.}
\label{tab:evidence}
\centering\scriptsize
\renewcommand{\arraystretch}{1.08}
\setlength{\tabcolsep}{2pt}
\begin{tabularx}{\columnwidth}{@{}p{1.55cm}YY@{}}
\toprule
\textbf{Evidence class} & \textbf{Examples} & \textbf{What it can support} \\
\midrule
Policy and governance & Data-use agreement, declassification rule, purpose limitation & What disclosure is authorized and which institution accepts residual risk \\
Formal and analytical & DP accountant, cryptographic proof, non-interference argument & A bounded property under an explicit mathematical and corruption model \\
Configuration and attestation & Signed configuration, TEE attestation, access-control state & Which code, policy, and isolation settings were active \\
Empirical and adversarial & Leakage benchmark, red-team report, participation-inference test & Observed resistance to a specified attack and test distribution \\
Operational and temporal & Scheduler records, telemetry, key rotation, incident history & Whether assumptions continued to hold during the deployment period \\
Provenance and audit & Signed manifest, model lineage, selective replay, audit record & Which workflow and artifact version the claim and evidence refer to \\
\bottomrule
\end{tabularx}
\end{table}

\subsection{Evidence adequacy and confidence}
For each claim, the register should record what inference the evidence supports:
its \emph{relevance} to the stated guarantee and observer, its \emph{coverage} of
stages/channels/validity period, and its \emph{strength} (formal, empirical, or
procedural), along with assumptions, scope, date, responsible party, and known
blind spots. This should not collapse into a single score: proofs and empirical
tests provide different confidence, and partial evidence is useful only if the
residual uncertainty stays explicit.

\subsection{Assurance is a maintained property}
Scientific-AI workflows evolve after initial review. Models are adapted, agents
receive new tools, scheduler policies change, instruments are recalibrated, and
collaborators enter or leave. Any of these events can alter an observer's view or
the meaning of permitted disclosure. We therefore treat assurance as a lifecycle rather than a one-time
certification. \emph{Design} fixes assets, observers, channels, disclosure rules,
and intended guarantees before mechanisms are chosen. \emph{Pre-deployment} binds
claims to configurations and collects proofs and approvals. \emph{Operation}
monitors assumption-sensitive evidence such as access state, scheduler behavior,
agent traces, and side-channel indicators. \emph{Change} re-evaluates affected
claims, and \emph{retirement} keeps the minimum evidence needed for later
verification while revoking access and deleting disclosure-prone traces. A versioned register links these stages, so a reviewer can ask whether the
claim held throughout the period that results, releases, and autonomous actions
depended on it.

\section{Applying the Framework Across the Lifecycle}\label{sec:lifecycle}
The register illustrates one deployment; the same six questions locate where assurance
breaks across the lifecycle (Fig.~\ref{fig:assurance}).

\subsection{Data and distributed training}
Federated learning is orchestration, not a guarantee: gradients, deltas, and
participation remain observable unless additional controls are used~\cite{zhu2019}.
Secure aggregation, MPC, and trusted execution protect only the values inside
their protocol under a stated corruption model. DP bounds sensitivity to a
protected unit~\cite{abadi2016}, but element~(1) remains open at the institution
level. Amplification by decentralization can make the guarantee depend on an
observer's network position~\cite{cyffers2022}, a dependence multi-facility runs
inherit. At scale, long adaptive workflows need accountants coupled to workflow
and scheduler state, not per-stage bookkeeping.

\subsection{Adaptation, evaluation, and release}
\label{sec:release}
Institutional fine-tuning can teach sensitive capabilities. DP reduces
memorization but does not prevent disclosing what the objective deliberately
teaches~\cite{coffey2024dpstego}, and trusted environments do not decide whether outputs are
authorized. Release therefore needs information-flow control (IFC): labels on
datasets/adapters/checkpoints, explicit declassification at release and tool
boundaries, mediation of tool outputs before they re-enter context, and tiered
adversarial tests. The resulting \emph{tiered-release certificate} records the
(model, adapter) versions, tier, declassification rules, tests/results, and a
validity period.
Modular-model IFC shows non-interference can be imposed across domains~\cite{tiwari2024};
keeping labels intact through distillation is the open part. Watermarking and
provenance then bind model, policy, and deployment context to the certificate~\cite{xu2026xmark}.

\subsection{Autonomous operation}
\label{sec:autonomous}
Autonomous operation turns privacy into a continuous control problem. Agents
exchange plans, tool arguments, retrieved context, and persistent memories,
which encode research strategy more directly than a gradient. AgentLeak reports
inter-agent leakage far exceeding final-output leakage~\cite{elyagoubi2026}; rates are
domain-specific, but the lesson holds: internal channels are in scope.
Compositional attacks can infer a secret from individually innocuous messages~\cite{patil2025}.
Operationally, instrument inter-agent messages, tool arguments, retrieved
context, and persistent memory, and replace raw content with task-conditioned
abstractions enforced for named recipients. Measure leakage per channel/observer
on scientific-domain task sets, with acceptance stated as a bound on what an
observer can infer about the protected hypothesis.

\section{Research Priorities and Outlook}\label{sec:agenda}

Read as a checklist, the empty cells of the six-element claim, instantiated for
science, define the agenda. Each priority can be stated as an asset, a current
gap, a research question, and required evidence.

\textbf{Institution-level guarantees.} The asset is participation, strategy, or
capability. Eq.~\ref{eq:pufferfish} states the target; what remains open is
choosing $\Theta$ for a scientific consortium and obtaining mechanisms that meet
the bound over a realistic $V_O$. The evidence that would populate C3 is
concrete: a leave-one-institution-out shadow-consortium attack reporting attacker
advantage over chance; a timing-only variant in which the observer sees per-round
completion times alone; and a combined attack over released weights, public
membership, and timing, which is exactly C3's observer set. An acceptance
threshold takes the form of advantage at most $\delta$ for a stated observer and
observation window. R\'enyi Pufferfish makes the underlying framework more tractable for
iterative learning~\cite{pierquin2024pufferfish}.

\textbf{Agent-communication privacy.} The assets are unpublished hypotheses,
target regions, and facility capabilities, which can leak through prompts, tool
calls, and message metadata. Section~\ref{sec:autonomous} states what a benchmark
must instrument; missing are domain task sets and acceptance thresholds for what
an agent may reveal to different observers.

\textbf{Cross-tier information flow.} The asset is tier-restricted knowledge.
Section~\ref{sec:release} gives the recipe and certificate; missing is a
guarantee that labels survive distillation.

\textbf{Privacy-compatible reproducibility.} The asset is protected workflow
context; the gap is that indiscriminate provenance disclosure can violate the
claim it is meant to evidence. Reproducibility needs enough evidence for a named
audience to verify claims and rerun permitted computations, not a public trace:
an auditor inspects a protected trace in a trusted environment, a reviewer
receives a commitment or proof, a future collaborator gets a policy-filtered
replay package. What is missing is an integrated design for FMs and agent
workflows, evaluated for scientific fidelity as well as secrecy.

\textbf{Leadership-scale accounting.} The asset and observer may change across a
long adaptive workflow; the gap is per-stage bookkeeping. Accounting must span
workflow and scheduler state, allocate loss across institutions, and report
utility, time, communication, energy, and privacy together, validated under the
deployed threat model~\cite{cebere2025}.

\textbf{Instrument side channels.} The asset is research strategy or facility
capability; leakage can occur through correlated sensor/actuator and controller
telemetry. The gap is the lack of representative threat models and trace data.
Evidence should include datasets for controllers and facility services, with
defenses judged against control-loop latency and calibration
accuracy~\cite{weiss2024,gao2025}.

\textbf{Limitations.} The six-element form is a specification aid: it does not itself provide
privacy, nor does a written claim prove an implementation meets it. The composite
scenario is illustrative; validating deployments should report per-institution
privacy loss, observer participation-inference advantage, utility/runtime
overhead, and what telemetry each observer can access. Assets and neighboring
worlds are domain-dependent; measurements are not worst-case; unknown observers
or channels may exist; evidence artifacts can themselves leak; and composition of
heterogeneous formal guarantees remains unresolved. Integrity, availability,
poisoning, safety, fairness, and legal compliance are out of scope except where
they affect disclosure.

\textbf{Outlook:} Success means bounded, falsifiable privacy statements, not a generic label.
Targets include: aggregation that cannot inspect any institution's update;
open-tier observers with limited advantage in inferring restricted participation;
agent messages that expose only approved abstractions; and audits that verify
policy compliance without revealing private context. A claim register should link
each property to an evidence artifact and validity period. Achieving this will
require coordinated work across privacy, ML, security, cryptography, and
AI-for-science to build systems institutions can trust and deploy.

\section*{Acknowledgment}
This manuscript has been co-authored by UT-Battelle, LLC under Contract No.\ DE-AC05-00OR22725 with the U.S.\ Department of Energy. The publisher, by accepting the article for publication, acknowledges that the U.S.\ Government retains a non-exclusive, paid-up, irrevocable, worldwide license to publish or reproduce the published form of this manuscript, or allow others to do so, for U.S.\ Government purposes. The Department of Energy will provide public access to these results of federally sponsored research in accordance with the DOE Public Access Plan (\url{https://energy.gov/downloads/doe-public-access-plan}). This work was co-supported by the U.S.\ Department of Energy, Office of Science, Advanced Scientific Computing Research, under Contract DE-AC02-06CH11357.

\bibliographystyle{IEEEtran}
\bibliography{references_used}

@misc{genesis2025,
  author       = {{The White House}},
  title        = {Launching the Genesis Mission},
  howpublished = {Executive Order 14363},
  year         = {2025},
  month        = nov,
  note         = {Signed 24 November 2025}
}

@article{xu2026optimal,
  title     = {Optimal client sampling in federated learning with client-level heterogeneous differential privacy},
  author    = {Xu, Jiahao and Hu, Rui and Kotevska, Olivera},
  journal   = {IEEE Internet of Things Journal},
  year      = {2026},
  publisher = {IEEE}
}

@article{kotevska2025privacy,
  title     = {Privacy Preservation from High-Performance Computing to Autonomous Science [{I}ndustrial and {G}overnmental {A}ctivities]},
  author    = {Kotevska, Olivera},
  journal   = {IEEE Computational Intelligence Magazine},
  volume    = {20},
  number    = {3},
  pages     = {5--6},
  year      = {2025},
  publisher = {IEEE}
}

@techreport{nist800223,
  title       = {High-Performance Computing Security: Architecture, Threat Analysis, and Security Posture},
  author      = {{National Institute of Standards and Technology}},
  institution = {NIST},
  type        = {Special Publication},
  number      = {800-223},
  year        = {2024}
}

@article{zhu2019,
  author  = {Zhu, Ligeng and Liu, Zhijian and Han, Song},
  title   = {Deep Leakage from Gradients},
  journal = {Advances in Neural Information Processing Systems},
  volume  = {32},
  year    = {2019}
}

@article{DworkR14,
  author  = {Dwork, Cynthia and Roth, Aaron},
  title   = {The Algorithmic Foundations of Differential Privacy},
  journal = {Foundations and Trends in Theoretical Computer Science},
  volume  = {9},
  number  = {3-4},
  pages   = {211--407},
  year    = {2014}
}

@article{abadi2016,
  author  = {Abadi, Mart{\'i}n and Chu, Andy and Goodfellow, Ian and McMahan, H. Brendan and Mironov, Ilya and Talwar, Kunal and Zhang, Li},
  title   = {Deep Learning with Differential Privacy},
  journal = {Proceedings of the 2016 ACM SIGSAC Conference on Computer and Communications Security},
  pages   = {308--318},
  year    = {2016},
  doi     = {10.1145/2976749.2978318}
}

@inproceedings{bonawitz2017secureagg,
  author    = {Bonawitz, Keith and Ivanov, Vladimir and Kreuter, Ben and Marcedone, Antonio and McMahan, H. Brendan and Patel, Sarvar and Ramage, Daniel and Segal, Aaron and Seth, Karn},
  title     = {Practical Secure Aggregation for Privacy-Preserving Machine Learning},
  booktitle = {Proceedings of the 2017 ACM SIGSAC Conference on Computer and Communications Security (CCS '17)},
  year      = {2017},
  pages     = {1175--1191},
  publisher = {ACM},
  doi       = {10.1145/3133956.3133982}
}

@inproceedings{yao1982gc,
  author    = {Yao, Andrew Chi-Chih},
  title     = {Protocols for Secure Computations},
  booktitle = {23rd Annual Symposium on Foundations of Computer Science (FOCS 1982)},
  year      = {1982},
  pages     = {160--164},
  doi       = {10.1109/SFCS.1982.38}
}

@techreport{costan2016sgx,
  author      = {Costan, Victor and Devadas, Srinivas},
  title       = {Intel {SGX} Explained},
  institution = {Cryptology ePrint Archive, Report 2016/086},
  year        = {2016},
  url         = {https://eprint.iacr.org/2016/086}
}

@inproceedings{shokri2017,
  author  = {Shokri, Reza and Stronati, Marco and Song, Congzheng and Shmatikov, Vitaly},
  title   = {Membership Inference Attacks Against Machine Learning Models},
  journal = {2017 IEEE Symposium on Security and Privacy},
  pages   = {3--18},
  year    = {2017},
  doi     = {10.1109/SP.2017.41}
}

@article{carlini2021,
  author  = {Carlini, Nicholas and Tram{\`e}r, Florian and Wallace, Eric and Jagielski, Matthew and Herbert-Voss, Ariel and Lee, Katherine and Roberts, Adam and Brown, Tom and Song, Dawn and Erlingsson, {\'U}lfar and Oprea, Alina and Raffel, Colin},
  title   = {Extracting Training Data from Large Language Models},
  journal = {30th USENIX Security Symposium},
  pages   = {2633--2650},
  year    = {2021}
}

@article{maini2024llmdatasetinference,
  author  = {Maini, Pratyush and Jia, Hengrui and Papernot, Nicolas and Dziedzic, Adam},
  title   = {{LLM} Dataset Inference: Did you train on my dataset?},
  journal = {arXiv preprint arXiv:2406.06443},
  year    = {2024},
  doi     = {10.48550/arXiv.2406.06443}
}

@inproceedings{ganju2018propertyinference,
  author    = {Ganju, Karan and Wang, Qi and Yang, Wei and Gunter, Carl A. and Borisov, Nikita},
  title     = {Property Inference Attacks on Fully Connected Neural Networks using Permutation Invariant Representations},
  booktitle = {Proceedings of the 2018 ACM SIGSAC Conference on Computer and Communications Security (CCS '18)},
  pages     = {619--633},
  publisher = {ACM},
  year      = {2018},
  doi       = {10.1145/3243734.3243834}
}

@article{kifer2014pufferfish,
  author  = {Kifer, D. and Machanavajjhala, A.},
  title   = {Pufferfish: A framework for mathematical privacy definitions},
  journal = {ACM Transactions on Database Systems},
  volume  = {39},
  number  = {1},
  pages   = {1--36},
  year    = {2014}
}

@article{linddun,
  author  = {Deng, Mina and Wuyts, Kim and Scandariato, Riccardo and Preneel, Bart and Joosen, Wouter},
  title   = {A privacy threat analysis framework: supporting the elicitation and fulfillment of privacy requirements},
  journal = {Requirements Engineering},
  volume  = {16},
  number  = {1},
  pages   = {3--32},
  year    = {2011}
}

@book{stride,
  author    = {Shostack, Adam},
  title     = {Threat Modeling: Designing for Security},
  publisher = {Wiley},
  year      = {2014}
}

@inproceedings{gsn,
  author    = {Kelly, Tim and Weaver, Rob},
  title     = {The Goal Structuring Notation---a safety argument notation},
  booktitle = {Proc. DSN Workshop on Assurance Cases},
  year      = {2004}
}

@article{nissenbaum2004ci,
  author  = {Nissenbaum, Helen},
  title   = {Privacy as contextual integrity},
  journal = {Washington Law Review},
  volume  = {79},
  number  = {1},
  pages   = {119--158},
  year    = {2004}
}

@inproceedings{kotevska2026scalable,
  author    = {Kotevska, Olivera and Nguyen, Trong and Ferreira da Silva, Rafael and Engelmann, Christian and Balaprakash, Prasanna},
  title     = {Scalable Federated Learning for Scientific Foundation Models on Leadership-Class Systems},
  booktitle = {Proceedings of the 6th Workshop on Machine Learning and Systems (EuroMLSys '26)},
  pages     = {439--446},
  publisher = {Association for Computing Machinery},
  doi       = {10.1145/3805621.3807639},
  year      = {2026}
}

@inproceedings{cyffers2022,
  title     = {Privacy amplification by decentralization},
  author    = {Cyffers, E. and Bellet, A.},
  booktitle = {AISTATS},
  volume    = {151},
  series    = {PMLR},
  pages     = {5334--5353},
  year      = {2022}
}

@inproceedings{elmrini2024,
  title     = {Privacy attacks in decentralized learning},
  author    = {El Mrini, A. and Cyffers, E. and Bellet, A.},
  booktitle = {NeurIPS},
  year      = {2024}
}

@article{zhu2024cc,
  title   = {Confidential computing on {NVIDIA} {Hopper} {GPUs}: A performance benchmark study},
  author  = {Zhu, J. and Yin, H. and Deng, P. and Almeida, A. and Zhou, S.},
  journal = {arXiv:2409.03992},
  year    = {2024}
}

@inproceedings{cebere2025,
  title     = {Tighter privacy auditing of {DP-SGD} in the hidden state threat model},
  author    = {Cebere, T. and Bellet, A. and Papernot, N.},
  booktitle = {IEEE S\&P},
  year      = {2025}
}

@article{elyagoubi2026,
  author  = {El Yagoubi, Farida and Al Mallah, Ranwa and Badu-Marfo, Godwin},
  title   = {{AgentLeak}: A full-stack benchmark for privacy leakage in multi-agent {LLM} systems},
  journal = {arXiv preprint arXiv:2602.11510},
  year    = {2026}
}

@article{patil2025,
  author  = {Patil, Vaidehi and Stengel-Eskin, Elias and Bansal, Mohit},
  title   = {The sum leaks more than its parts: Compositional privacy risks and mitigations in multi-agent collaboration},
  journal = {arXiv preprint arXiv:2509.14284},
  year    = {2025}
}

@article{cebere2026zeroauditing,
  author  = {Cebere, Tudor and Even, Mathieu and Bleistein, Linus and Bellet, Aur{\'e}lien},
  title   = {Privacy Auditing with Zero (0) Training Run},
  journal = {arXiv preprint arXiv:2605.14591},
  year    = {2026}
}

@article{weiss2024,
  author  = {Weiss, Roy and Ayzenshteyn, Daniel and Amit, Guy and Mirsky, Yisroel},
  title   = {What Was Your Prompt? A Remote Keylogging Attack on {AI} Assistants},
  journal = {33rd USENIX Security Symposium (USENIX Security 24)},
  pages   = {3367--3384},
  year    = {2024},
  publisher = {USENIX Association}
}

@inproceedings{gao2025,
  author    = {Gao, Zihao and others},
  title     = {I know what you said: Unveiling hardware cache side-channels in local large language model inference},
  booktitle = {34th USENIX Security Symposium (USENIX Security 25)},
  pages     = {1649--1668},
  year      = {2025}
}

@inproceedings{dutta2023spygpu,
  author    = {Dutta, Sankha Baran and Naghibijouybari, Hoda and Gupta, Arjun and Abu-Ghazaleh, Nael and Marquez, Andres and Barker, Kevin},
  title     = {Spy in the {GPU}-box: Covert and Side Channel Attacks on Multi-{GPU} Systems},
  booktitle = {Proceedings of the 50th Annual International Symposium on Computer Architecture (ISCA)},
  year      = {2023}
}

@article{tiwari2024,
  author = {Tiwari, Trishita and Gururangan, Suchin and Guo, Chuan and Hua, Weizhe and Kariyappa, Sanjay and Gupta, Udit and Xiong, Wenjie and Maeng, Kiwan and Lee, Hsien-Hsin S. and Suh, G. Edward},
  title = {Information Flow Control in Machine Learning through Modular Model Architecture},
  journal = {33rd USENIX Security Symposium (USENIX Security 24)},
  pages = {6921--6938},
  year = {2024},
  publisher = {USENIX Association}
}

@inproceedings{xu2026xmark,
  title={XMark: Reliable Multi-Bit Watermarking for LLM-Generated Texts},
  author={Xu, Jiahao and Hu, Rui and Kotevska, Olivera and Zhang, Zikai},
  booktitle={Proceedings of the 64th Annual Meeting of the Association for Computational Linguistics (Volume 1: Long Papers)},
  pages={14747--14763},
  year={2026}
}

@ARTICLE{kairouz2021,
  author = {Peter Kairouz and H. Brendan McMahan and Brendan Avent and
  Aurélien Bellet and others},
  title = {{A}dvances and {O}pen {P}roblems in {F}ederated {L}earning},
  journal = {Foundations and Trends® in Machine Learning},
  year = {2021},
  volume = {14},
  number = {1--2},
  pages = {1--210}
}

@article{li2026scalable,
  author  = {Li, Yijiang and Li, Zilinghan and Chard, Kyle and Foster, Ian and Munson, Todd and Madduri, Ravi and Kim, Kibaek},
  title   = {Scalable cross-facility federated learning for scientific foundation models on multiple supercomputers},
  journal = {arXiv preprint arXiv:2603.19544},
  year    = {2026}
}

@inproceedings{kotevska2026dptwolevel,
  author    = {Kotevska, Olivera and Patton, Robert and Jha, Sumit and Balaprakash, Prasanna},
  title     = {{DP-TwoLevel}: Two-Stage Gradient Subspace Learning for Differentially Private Federated Learning},
  booktitle = {SPIE Conference on Assurance and Security for AI-enabled Systems},
  year      = {2026}
}

@inproceedings{kim-kotevska2025privacy,
  author    = {Kim, Kibaek and Raghavan, Krishnan and Kotevska, Olivera and Dorier, Matthieu and Madduri, Ravi and Ryu, Minseok and Yousefian, Farzad},
  title     = {Privacy-preserving federated learning for science: Challenges and research directions},
  booktitle = {IEEE International Conference on Big Data (BigData)},
  year      = {2025}
}

@inproceedings{coffey2024dpstego,
  author    = {Coffey, Sean M. and Catudal, Joseph W. and Bastian, Nathaniel D.},
  title     = {Differential privacy to mathematically secure fine-tuned large language models for linguistic steganography},
  booktitle = {Assurance and Security for AI-enabled Systems},
  series    = {Proc. SPIE},
  volume    = {13054},
  pages     = {130540K},
  year      = {2024},
  doi       = {10.1117/12.3013129}
}

@inproceedings{pierquin2024pufferfish,
  author    = {Pierquin, Cl{\'e}ment and Bellet, Aur{\'e}lien and Tommasi, Marc and Boussard, Matthieu},
  title     = {R{\'e}nyi Pufferfish Privacy: General Additive Noise Mechanisms and Privacy Amplification by Iteration via Shift Reduction Lemmas},
  booktitle = {International Conference on Machine Learning (ICML)},
  year      = {2024}
}
\end{document}